\documentclass[aps,preprint]{revtex4}%
\usepackage{amsfonts}
\usepackage{amsmath}
\usepackage{amssymb}
\usepackage{graphicx}%
\begin{document}
\title[Cooper pairing and superconductivity on a sphere]{Cooper pairing and superconductivity on a spherical surface: applying the
Richardson model to a multielectron bubble in liquid helium}
\author{J. Tempere$^{1,2}$, V.N. Gladilin$^{1,\ast}$, I.F. Silvera$^{2}$, J.T.
Devreese$^{1}$}
\affiliation{$^{1}$TFVS, Departement Fysica, Universiteit Antwerpen, Universiteitsplein 1,
B2610 Antwerpen, Belgium}
\affiliation{$^{2}$Lyman Laboratory of Physics, Harvard University, Cambridge MA 02138}


\begin{abstract}
Electrons in a multielectron bubble in helium form a spherical,
two-dimensional system coupled to the ripplons at the bubble surface. The
electron-ripplon coupling, known to lead to polaronic effects, is shown to
give rise also to Cooper pairing. A Bardeen-Cooper-Schrieffer (BCS)
Hamiltonian arises from the analysis of the electron-ripplon interaction in
the bubble, and values of the coupling strength are obtained for different
bubble configurations. The BCS Hamiltonian on the sphere is analysed using the
Richardson method. We find that although the typical ripplon energies are
smaller than the splitting between electronic levels, a redistribution of the
electron density over the electronic levels is energetically favourable as
pairing correlations can be enhanced. The density of states of the system with
pairing correlations is derived. No gap is present, but the density of states
reveals a strong step-like increase at the pair-breaking energy. This feature
of the density of states should enable the unambiguous detection of the
proposed state with pairing correlations in the bubble, through either
capacitance spectroscopy or tunneling experiments, and allow to map out the
phase diagram of the electronic system in the bubble.

\end{abstract}
\date{}
\maketitle

\tighten

\section{Introduction}

\bigskip

Spherical shells of charge carriers appear in a multitude of systems, such as
multielectron bubbles in liquid helium \cite{VolodinJETP26}, metal nanoshells
coating a non-conducting nanograin \cite{AverittPRL78}, carbon cages and
fullerenes. Although the properties of flat two-dimensional systems have been
widely studied, revealing new physics, the properties of \emph{spherical}
two-dimensional systems are much less well-studied. In this paper, we
investigate the possibility and the properties of Cooper pairing in the
spherical geometry.

The particular spherical two-dimensional system that we focus on is the
multielectron bubble in liquid helium. When a flat surface of helium is
charged with electrons above a critical charge density, an instability occurs
with the surface opening to subsume a large number of electrons forming a
bubble. These multielectron bubbles (MEBs) are typically micron-sized cavities
inside liquid helium, containing a nanometer thin film of electrons on the
inner surface of the bubble. The cavity is forced open by the Coulomb
repulsion of the electrons which is balanced by the surface tension of the
helium. The equilibrium shape of the bubble is spherical, with a radius $R$
determined by the number of electrons and the pressure on the helium.

The bare single electron states on the surface of the spherical bubble are
angular momentum eigenstates and have discrete energies, characterized by the
angular momentum $L$ and with degeneracy $2L+1$. At low temperature there will
be a well defined Fermi surface located at the highest occupied state.
Small-amplitude shape oscillations, including surface waves, can be quantized
as spherical ripplons. The electrons can interact with these ripplons, and we
will show that this leads to an attractive effective interaction between electrons.

This paper has two distinct but interwoven parts. In the first part (section
II) we discuss in detail how the interactions between electrons and ripplons
in the bubble can lead to a Cooper pairing scenario. The goal of the first
part is to show that a BCS-type Hamiltonian provides a plausible description
of the electronic system in multielectron bubbles at low temperatures, and to
illustrate the relevant values of the parameters of the model Hamiltonian. In
the second part (sections III-IV) we investigate the general properties of
this Hamiltonian using the Richardson solution for the reduced
BCS\ Hamiltonian describing pairing e.g. in nanograins. Both the ground state
properties and the density of states of a spherical two-dimensional BCS
system, are derived and discussed.

\bigskip

\section{Cooper pairs on a spherical surface}

First, we investigate the three ingredients of the full Hamiltonian: the
electronic part, the ripplonic part, and the electron-ripplon coupling. Then,
we analyze the effective interaction between electrons, resulting from both
the Coulomb interactions and the ripplon-mediated electron-electron
interaction. Finally, we arrive at a BCS-type Hamiltonian for the
multielectron bubble.

\bigskip

\subsection{Electrons and ripplons in the bubble}

\bigskip

\textbf{The spherical 2D electron system -- }The well-known Hamiltonian of
interacting electrons in a flat 2D electron gas (2DEG) in the jellium model
can be written in second quantization as
\begin{align}
\hat{H}_{\text{e}}^{\text{flat}}  &  =\sum_{\mathbf{k},\sigma}\epsilon_{k}%
\hat{c}_{\mathbf{k},\sigma}^{\dag}\hat{c}_{\mathbf{k},\sigma}\nonumber\\
&  +\sum_{\mathbf{q}>0}\sum_{\mathbf{k},\sigma}\sum_{\mathbf{k}^{\prime
},\sigma^{\prime}}v_{q}\hat{c}_{\mathbf{k}+\mathbf{q},\sigma}^{\dag}\hat
{c}_{\mathbf{k}^{\prime}-\mathbf{q},\sigma^{\prime}}^{\dag}\hat{c}%
_{\mathbf{k}^{\prime},\sigma^{\prime}}\hat{c}_{\mathbf{k},\sigma}.
\label{Helflat}%
\end{align}
where $\hat{c}_{\mathbf{k},\sigma}^{\dag}\hat{c}_{\mathbf{k},\sigma}$ create
and destroy an electron with wave number $\mathbf{k}$ and spin $\sigma$, and
\begin{equation}
\epsilon_{k}=\frac{\hbar^{2}}{2m_{\text{e}}}k^{2},\text{ }v_{k}=\frac{e^{2}%
}{2\varepsilon A}\frac{1}{k}%
\end{equation}
where $m_{\text{e}}$ is the electron mass, $A$ is the surface of the 2D
system, $\varepsilon$ is the permittivity of the medium and $e$ is the
electron charge.

The Hamiltonian of the interacting \emph{spherical} electronic system has a
very similar form, provided that one uses spherical harmonics $Y_{L,m}%
(\theta,\phi)$ instead of plane waves as the single-particle basis functions.
For this purpose we use the operators $\hat{c}_{L,m}^{\dag}$ and $\hat
{c}_{L,m}$ that create, resp., annihilate an electron in the angular momentum
eigenstate $(L,m)$, i.e. $\psi_{L,m}(\theta,\phi)=Y_{L,m}(\theta,\phi)$. The
Hamiltonian of the interacting spherical two-dimensional electron gas (S2DEG)
becomes
\begin{align}
\hat{H}_{\text{e}}^{\text{sphere}}  &  =\sum_{L,m,\sigma}\epsilon_{L}\hat
{c}_{L,m,\sigma}^{\dag}\hat{c}_{L,m,\sigma}\nonumber\\
&  +\sum_{J>0,n}\sum_{L,m,\sigma}\sum_{L^{\prime},m^{\prime},\sigma}v_{L}%
\hat{c}_{(L,m)\otimes(J,n),\sigma}^{\dag}\hat{c}_{(L^{\prime},m^{\prime
})\otimes(J,-n),\sigma^{\prime}}^{\dag}\hat{c}_{L^{\prime},m^{\prime}%
,\sigma^{\prime}}\hat{c}_{L,m,\sigma} \label{Hel}%
\end{align}
where we use the notation $\sum_{L,m}=\sum_{L=0}^{\infty}\sum_{m=-L}^{L}$ and
\begin{equation}
\epsilon_{L}=\frac{\hbar^{2}}{2m_{\text{e}}}\frac{L(L+1)}{R^{2}}\text{, }%
v_{L}=\dfrac{e^{2}}{2\varepsilon R}\dfrac{(-1)^{L}}{2L+1},
\end{equation}
where $R$ is the radius of the sphere. The typical scale of the kinetic energy
$\epsilon_{1}=\hbar^{2}/(m_{\text{e}}R^{2})$ is listed for several bubble
sizes and pressures in table I. The operator $\hat{c}_{(L,m)\otimes
(J,n),\sigma}^{+}$ creates a spin $\sigma$ electron in a single particle state
resulting from adding the angular momenta $(L,m)$ and $(J,n).$ Formally, we
have%
\begin{equation}%
\begin{array}
[c]{c}%
\hat{c}_{(L,m)\otimes(J,n),\sigma}^{+}=%
{\displaystyle\sum\limits_{L^{\prime}=|L-J|}^{L+J}}
{\displaystyle\sum\limits_{m^{\prime}=-L^{\prime}}^{L^{\prime}}}
\sqrt{\dfrac{(2L+1)(2J+1)}{4\pi(2L^{\prime}+1)}}\left\langle J,0;L,0|L^{\prime
},0\right\rangle \\
\times\text{ }\left\langle J,n;L,m|L^{\prime},m^{\prime}\right\rangle \text{
}\hat{c}_{L^{\prime}m^{\prime},\sigma}^{+},
\end{array}
\end{equation}
where $\left\langle J,n;L,m|L^{\prime},m^{\prime}\right\rangle $ is the
Clebsch-Gordan coefficient for combining the angular momenta $(L,m)$ and
$(J,n)$ into a state of angular momentum $(L^{\prime},m^{\prime})$. The role
of the momentum is now taken by the angular momentum. Indeed, taking
$L/R\rightarrow k$ for large $L$ links the result for the spherical case to
that for the flat case. This link was noted previously for the structure
factor of the spherical 2D electron gas \cite{TemperePRB65}, and its response
to a weak magnetic field.

When the Coulomb energy is small compared to the kinetic energy, the electrons
fill up a Fermi sea of angular momentum states $L$ (with degeneracy $2L+1$) up
to a Fermi level $L=L_{F}$. The level splitting at the Fermi level,
$\Delta\epsilon=\epsilon_{L_{F}}-\epsilon_{L_{F}-1}$, is given in Table I for
some typical bubbles. The typical value of the distance between electronic
levels is THz (or mK).

Note that the $J=0$ term is absent in the Coulomb part of the Hamiltonian
(\ref{Hel}): this term is exactly cancelled by the surface tension energy of
the helium as shown in Ref. \cite{TemperePHYSE22}. In Hamiltonian
(\ref{Helflat}) for the flat 2D electron gas in jellium, the $\mathbf{q}=0$
term is absent because it is canceled by a homogeneous positive background
introduced in the jellium model. Thus the surface tension energy in MEBs takes
a role similar to the homogenous positive background in jellium.

\bigskip

\begin{table}[ptb]
\caption{Several typical length and energy scales for the electron-ripplon
system in the bubble are given in this table, for bubbles with different
numbers of electrons and subjected to different pressures. The first row lists
the bubble radius in microns. The second row lists the kinetic energy scale
$\epsilon_{1}=\hbar^{2}/(m_{\text{e}}R^{2})$ of electrons in the bubble, the
energy of angular momentum level $L$ being $\epsilon_{L}=\epsilon_{1}%
L(L+1)/2$. The third row gives the electronic level splitting $\Delta
\epsilon=\epsilon_{1}(L_{F}+1)$, at the Fermi level, between the subsequent
angular momentum levels of the spherical 2D electron gas. The fourth row lists
the energy scale for ripplons, $\hbar\omega_{r}=\hbar\lbrack\sigma/(\rho
R^{3})]^{1/2}$ in $\mu$K (1 $\mu$K corresponds to 0.1309 MHz). The fifth row
gives the strength of the electric field at the bubble surface, pressing the
electrons to the helium surface and resulting in electron-ripplon coupling.
The sixth row provides values for the electron-ripplon coupling constant $g$
in expression (\ref{Melrip}).}%
\label{tab1}
\begin{ruledtabular}
\begin{tabular}{l|ccc|ccc|ccc}
&  & $N=10^{4}$ &  &  & $N=10^{5}$ &  &  & $N=10^{6}$ &  \\
$p$ (Pa) & 0 & 10$^{2}$ & 10$^{4}$ & 0 & 10$^{2}$ & 10$^{4}$ & 0 & 10$^{2}$
& 10$^{4}$ \\ \hline\hline
$R$ ($\mu $m) & 1.062 & .5236 & .1709 & 4.937 & 1.6930 & .5409 & 22.93 &
5.393 & 1.711 \\
$\epsilon _{1}$ (mK) & .7836 & 3.225 & 30.27 & .0363 & .3084 & 3.022 &
1.68$\times$10$^{-3}$ & .0304 & .3021 \\
$\Delta \epsilon $ (mK) & 55.64 & 229.0 & 2149. & 8.127 & 69.10 & 673.0 &
1.191 & 21.53 & 213.9 \\
$\hbar \omega _{r}$ ($\mu $K) & 10.99 & 31.76 & 170.3 & 1.097 & 5.463 & 30.25
& .1096 & .9611 & 5.378 \\
$\mathbf{E}$ (kV/cm) & 63.80 & 262.6 & 2465. & 29.54 & 251.2 & 2461. & 13.70
& 247.6 & 2459. \\
$g$ (mK) & 25.87 & 124.1 & 1909. & 4.933 & 49.70 & 803.3 & .9813 & 20.61 &
338.5
\end{tabular}
\end{ruledtabular}
\end{table}

\bigskip

\textbf{Ripplons on the bubble surface} -- Small-amplitude oscillations of the
bubble surface can be quantized, leading to the concept of spherical ripplons.
The ripplon gas (excluding the breathing mode) is described by the
Hamiltonian
\begin{equation}
\hat{H}_{\text{ripl}}=\sum_{L>0,m}\hbar\omega_{L}\hat{a}_{L,m}^{+}\hat
{a}_{L,m}. \label{Hrip}%
\end{equation}
The bare ripplonic frequencies for this system, at a pressure $p$, are
\cite{TemperePRL87}
\begin{equation}
\omega_{L}=\sqrt{\frac{\sigma}{\rho R^{3}}(L+1)(L^{2}+L+2)+\frac{p}{\rho
R^{2}}2(L+1)},
\end{equation}
where $\sigma=3.6\times10^{-4}$ J/m$^{2}$ is the surface tension of helium,
and $\rho=145$ kg/m$^{3}$ is its density. In the surface tension dominated
regime ($pR/\sigma<1$), $\omega_{L}=\omega_{r}L^{3/2}$ with $\omega
_{r}=[\sigma/(\rho R^{3})]^{1/2}$. Typical ripplon frequencies lie in the
MHz-GHz (or $\mu$K) range. The ripplon Green's function is defined by
\begin{align*}
D(L,m;t)  &  =-i\left\langle \mathcal{T}[\hat{A}_{L,m}(t)\hat{A}_{L,m}%
^{+}(0)]\right\rangle \\
&  \text{with }\hat{A}_{L,m}=\hat{a}_{L,m}+\hat{a}_{L,-m}^{+},
\end{align*}
where $\mathcal{T}$ is the time ordering operator and $\hat{A}_{L,m}$ is a sum
of ripplon creation and annihilation operators. The unperturbed ripplon
propagator (corresponding to a system described by $\hat{H}_{\text{ripl}}$
above) in the frequency domain is
\[
D^{(0)}(L,m;\omega)=\dfrac{2\hbar\omega_{L}}{(\hbar\omega)^{2}-(\hbar
\omega_{L})^{2}+i\eta}.
\]
where $\eta$ is a positive infinitesimal.

\bigskip

\textbf{Electron-ripplon interaction} -- The electron-ripplon interaction can
be written as
\begin{equation}
\hat{H}_{\text{int}}=\sum_{J,n}M_{J}\hat{A}_{J,n}\sum_{L,m,\sigma}\hat
{c}_{(L,m)\otimes(J,n),\sigma}^{+}\hat{c}_{L,m,\sigma}, \label{Helrip}%
\end{equation}
where $M_{J}$ is the electron-ripplon interaction amplitude and $\sum
_{L,m,\sigma}\hat{c}_{(L,m)\otimes(J,n),\sigma}^{+}\hat{c}_{L,m,\sigma}$ is
the $(J,n)$ spherical component of the electron density.

The interaction between the electrons and the ripplons comes about due to the
presence of an electric field, generated by the electrons themselves and
pressing the electrons against the helium surface. This is the electric
pressing field $E=eN/(2R^{2})$, directed radially. When a ripplon is present,
it moves the electrons in the electric field generated by all other electrons
and this results in an interaction energy. The interaction energy is the
product of the displacement caused by the ripplon and the electric field,
summed for all electrons, similarly as in \cite{KliminSSC126}. Rewriting this
interaction energy in second quantization operators, we find the interaction
Hamiltonian (\ref{Helrip}) with the interaction amplitude%

\begin{equation}
M_{J}=g\frac{(J+1/2)^{1/2}}{[(J+1)(J^{2}+J+2+2pR/\sigma)]^{1/4}},
\label{Melrip}%
\end{equation}
where the coupling constant due to the pressing electric field $E$ (see Table
I) is given by%
\begin{equation}
g^{\text{(e)}}=\frac{1}{2\sqrt{\pi}}\frac{-eE}{R}\sqrt{\frac{\hbar}{2\rho
R\omega_{r}}}.
\end{equation}
An additional contribution to the interaction energy between electrons and
ripplons can be derived as the change in polarization energy of the
electron-helium system when the helium surface is deformed and the electrons
are at rest. This mechanism for coupling was first derived by M. Cole
\cite{ColePRB2} for electrons on a flat surface. Following the arguments of
Cole, we obtain a similar expression for the interaction amplitude
(\ref{Melrip}) but with a different coupling constant
\begin{equation}
g^{\text{(p)}}=\frac{9\sqrt{\pi}}{8}\frac{\varepsilon-1}{\varepsilon+1}%
\frac{e^{2}}{4\pi d^{2}R}\sqrt{\frac{\hbar}{2\rho R\omega_{r}}},
\end{equation}
where $d$ is the expectation value for the distance between the electron and
the helium surface. The total electron-ripplon coupling constant is then
$g=g^{\text{(p)}}+g^{\text{(e)}}.$

A single electron coupled to a bath of ripplons forms a ripplonic polaron
\cite{JacksonPRB24}. In a multielectron bubble, the electric field pressing
the electrons against the helium surface can be much larger than the field
achievable on a flat helium surface, so that the ripplonic polarons will be in
the strong coupling regime, and can even form a Wigner lattice of ripplonic
polarons \cite{TempereEPJB32}.

\bigskip

\subsection{Effective electron-electron interaction}

\bigskip

\begin{figure}
[ptb]
\begin{center}
\includegraphics[
height=2.5513in,
width=3.0676in
]%
{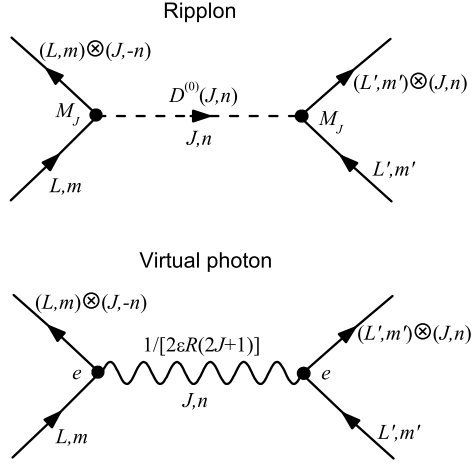}%
\caption{Electrons interact through the exchange of virtual photons or
ripplons. The vertex contributions and propagator for both interactions are
shown in this figure.}%
\label{fig1}%
\end{center}
\end{figure}

\textbf{Cooper's argument} -- In this subsection, we follow Cooper's argument
for pairing \cite{CooperPR104} and apply this to the present case of electrons
and ripplons in the bubble. The effective electron-electron interaction is the
sum of the Coulomb interaction between the electrons and a ripplon-mediated
attractive interaction between the electrons. In \cite{TemperePRB65} we drew a
Feynman diagram to represent the Coulomb interaction between the electrons.
Now, we can add another diagram with the same electron propagator lines, but
instead of exchanging a virtual photon, exchanging a virtual ripplon. This is
shown in Fig. \ref{fig1}.%

The two diagrams in Fig. \ref{fig1} can be represented by a single diagram
using an effective interaction \
\begin{equation}
V_{\text{eff}}(L,m;\omega)=\dfrac{e^{2}}{2\varepsilon R}\dfrac{1}{2L+1}%
+M_{L}^{2}\dfrac{2\hbar\omega_{L}}{(\hbar\omega)^{2}-(\hbar\omega_{L}%
)^{2}+i\eta}.
\end{equation}
Both the Coulomb and the ripplon-exchange interaction are given by a product
of two vertex factors and one virtual particle propagator. The total
electron-electron interaction Hamiltonian can be written as
\begin{align*}
\hat{H}_{\text{int}}(t)  &  =\sum_{L,m,\sigma}\sum_{L^{\prime},m^{\prime
},\sigma^{\prime}}\sum_{J,n}\left[  \int\frac{d\omega}{2\pi}V_{\text{eff}%
}(J,n;\omega)e^{i\omega t}\right] \\
&  \times(-1)^{n}\hat{c}_{(L,m)\otimes(J,n),\sigma}^{+}\hat{c}_{(L^{\prime
},m^{\prime})\otimes(J,-n),\sigma^{\prime}}^{+}\hat{c}_{L^{\prime},m^{\prime
},\sigma^{\prime}}\hat{c}_{L,m,\sigma}.
\end{align*}
Let's study for which regimes $V_{\text{eff}}(J,n;\omega)$ is attractive, i.e.
for which values of $(J,n;\omega)$ the ripplonic part dominates and is
attractive. It is clear that small energy transfers make the ripplonic part
attractive because $D^{(0)}(L,\omega\rightarrow0)=-2/(\hbar\omega_{L})$.
Moreover, in general ripplon exchange will indeed occur with $\omega=0$. The
reason for this is that the ripplonic energies are much smaller than the
electronic level spacing, as can be seen from comparing rows 3 and 4 of table
I. For ripplons with $L$ smaller than $\sim10^{3}$, the absorption or emission
of a ripplon with angular momentum $L$ cannot change the angular momentum of
the electron due to energy conservation requirements. The effective
interaction at $\omega=0$ is
\begin{equation}
V_{\text{eff}}(L,m;0)=\dfrac{e^{2}}{2\varepsilon R}\dfrac{1}{2L+1}%
-\dfrac{2M_{L}^{2}}{\hbar\omega_{L}}.
\end{equation}
The attractive interaction dominates strongly at small $L,$ since
$2g^{2}/(\hbar\omega_{r})\gg e^{2}/2\varepsilon R$ as can be checked for
typical bubbles by substituting the values from Table I$.$ It is strongest for
small $L$ and decreases roughly as $L^{-2}$:
\begin{equation}
V_{\text{eff}}(L,m;0)\approx-\dfrac{2M_{L}^{2}}{\hbar\omega_{L}}=-\frac
{2g^{2}}{\hbar\omega_{r}}\frac{(L+1/2)}{(L+1)(L^{2}+L+2+2pR/\sigma)}.
\end{equation}
The strength of the effective interaction is $2g^{2}/(\hbar\omega_{\sigma})$.
Since the $g$ is of the order of mK and $\hbar\omega_{r}$ of the order of
$\mu$K (see Table I), the effective attractive interaction is large as
compared the relevant energy scales $g$, $\hbar\omega_{r}$, and $\epsilon_{1}%
$. We are clearly in a strong coupling regime, in agreement with the results
from \cite{TempereEPJB32}.

\bigskip

\textbf{Intralevel pairing} -- Since the ripplon energies are much smaller
than the electronic level spacing at the Fermi level, the attractive
interaction only takes place between two electrons on the same angular
momentum level, and these electrons will be scattered into final states also
on that angular momentum level. An electron in angular momentum state
$\left\vert \text{initial}\right\rangle =\left\vert L,m\right\rangle $, which
emits a spherical ripplon in angular momentum state $\left\vert
J,n\right\rangle $, finds itself in the following superposition of angular
momentum states
\begin{align*}
\left\vert \text{final}\right\rangle  &  =\sum_{L^{\prime}=|L-J|}^{L+J}%
\sqrt{\dfrac{(2L+1)(2J+1)}{4\pi(2L^{\prime}+1)}}\left\langle L,0;J,0|L^{\prime
},0\right\rangle \\
&  \times\left\langle L,m;J,-n|L^{\prime},m-n\right\rangle \text{ }\left\vert
L^{\prime},m-n\right\rangle .
\end{align*}
The projection of this final state $\left\vert \text{final}\right\rangle $ on
the angular momentum level $L$ of the initial state is
\begin{align*}
f_{CG}[(L,m),(J,-n)]  &  =\sqrt{\dfrac{2J+1}{4\pi}}\left\langle
L,0;J,0|L,0\right\rangle \\
&  \times\left\langle L,m;J,-n|L,m-n\right\rangle .
\end{align*}
Thus, the scattering process between two electrons with spin $\sigma$ and
$\sigma^{\prime}$ on the angular momentum level $L$ can be described in second
quantization as
\begin{align*}
\hat{H}_{\text{int,}L}  &  =\sum_{m=-L}^{L}\sum_{m^{\prime}=-L}^{L}\sum
_{J,n}f_{CG}[(L,m),(J,-n)]f_{CG}[(L,m^{\prime}),(J,n)]\\
&  \times V_{\text{eff}}(J,n;0)\hat{c}_{L,m-n,\sigma}^{+}\hat{c}_{L,m^{\prime
}+n,\sigma^{\prime}}^{+}\hat{c}_{L,m^{\prime},\sigma^{\prime}}\hat
{c}_{L,m,\sigma}%
\end{align*}
This interaction Hamiltonian derived for the multielectron bubble is already
close to a BCS-like interaction Hamiltonian. It involves only electrons on the
same angular momentum level and couples them with an attractive potential.

An initial state with a pair characterized by $\{m,m^{\prime}\}$ can be
scattered into a pair with $\{m-n,m^{\prime}+n\}$ in various ways: namely by
the scattering of a virtual ripplon with $J=n,n+1,n+2,....$ provided that $J$
is an even mode to obey the triangle rule of addition of angular momenta. All
these processes are indistinguishable (initial and final states are exactly
the same) and their diagrams should be added to get the overall amplitude.
Different combinations of Clebsch-Gordan coefficients will occur, which can
take either a positive or a negative sign: the different diagrams can
interfere constructively but also destructively. The only case where we are
sure that the different contributions will interfere constructively, is for
pairs of electrons with opposite angular momentum $m^{\prime}=-m$. The reason
for this is that $\left\langle L,m;J,n|L,m+n\right\rangle $ has the same sign
as $\left\langle L,-m;J,-n|L,-m-n\right\rangle $. We investigated this point
numerically, and found that the total effective interaction is indeed strongly
reduced for pairs that do not have opposite angular momentum ($m^{\prime}%
\neq-m$). However, for $m^{\prime}=m$ we find that the effective interaction
potential is not reduced and is only weakly dependent on $m$.

\bigskip

\begin{figure}
[ptb]
\begin{center}
\includegraphics[
height=2.5786in,
width=3.4985in
]%
{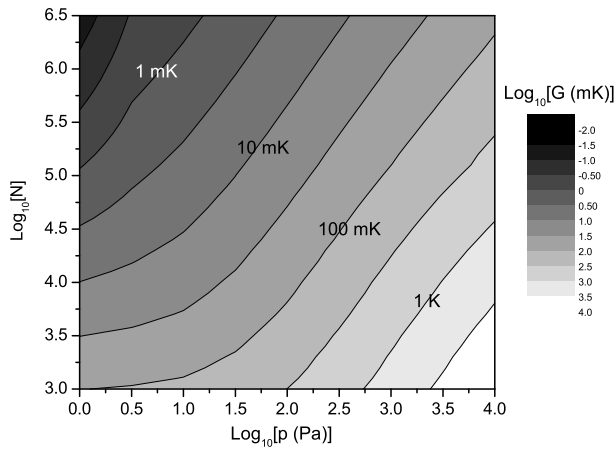}%
\caption{BCS interaction strength $G$ (in mK) as a function of pressure $p$
and number of electrons $N$ in the bubble.}%
\label{fig2}%
\end{center}
\end{figure}

\textbf{Effective Hamiltonian }-- From the previous section we know that the
interaction between the electrons can be written in the form of a BCS
interaction Hamiltonian given by%

\begin{equation}
\hat{H}_{\text{int}}=-\sum_{L}\sum_{m=-L}^{L}\sum_{m^{\prime}=-L}^{L}\tilde
{V}_{m,m^{\prime},L}\text{ }\hat{c}_{L,-m^{\prime}\downarrow}^{+}\hat
{c}_{L,m^{\prime};\uparrow}^{+}\hat{c}_{L,m;\uparrow}\hat{c}_{L,-m;\downarrow
}, \label{HBCS}%
\end{equation}
where $\sigma=$ $\uparrow,\downarrow$ denotes spin up and spin down and
\begin{align}
\tilde{V}_{m,m^{\prime},L}  &  =\sum_{J=\max[2,|m-m^{\prime}|]}^{2L}%
\frac{2g^{2}}{\hbar\omega_{\sigma}}\frac{(J+1/2)}{(J+1)(J^{2}+J+2+2pR/\sigma
)}\nonumber\\
&  \times f_{CG}[(L,m),(J,m^{\prime}-m)]f_{CG}[(L,-m),(J,m-m^{\prime})].
\label{Vmm}%
\end{align}
Due to the coefficients $\left\langle L,0;J,0|L,0\right\rangle $ the summation
cannot run further than $2L$ and only even values of $J$ contribute. The
summation starts from $n=|m-m^{\prime}|$, or if this is less than 2, it starts
at $J=2$. The $J=0$ deformation is not taken into account (it is the radius of
the bubble), and the $J=1$ deformation is a uniform translation which cannot
couple to the internal degrees of freedom. To proceed, we will introduce an
averaged interaction amplitude at the Fermi angular momentum level:
\begin{equation}
G=\frac{1}{(2L_{F}+1)^{2}}\sum_{m,m^{\prime}}\tilde{V}_{m,m^{\prime},L_{F}}.
\end{equation}
The interaction amplitude still depends on the angular momentum level $L$.
But, as we shall see in the next section, pair correlations occur only in
levels close to $L_{F}$ where the $L$-dependence of the interaction amplitude
can be neglected. So we use the value of the interaction amplitude at $L_{F}$
also for levels close to $L_{F}$. Values for $G$ for various configurations
are given in Figure \ref{fig2}.%

With this, we have established that the properties of the ripplon-mediated
electron-electron interaction lead to a Cooper-type attractive interaction
between the electrons, and to a BCS-like Hamiltonian%

\begin{align}
\hat{H}_{\text{eff}}  &  =\sum_{L,m,\sigma}\epsilon_{L}\hat{c}_{L,m;\sigma
}^{+}\hat{c}_{L,m;\sigma}\nonumber\\
&  -G\sum_{L}\sum_{m,m^{\prime}=-L}^{L}\hat{c}_{L,-m^{\prime}\downarrow}%
^{+}\hat{c}_{L,m^{\prime};\uparrow}^{+}\hat{c}_{L,m;\uparrow}\hat
{c}_{L,-m;\downarrow}. \label{Heff}%
\end{align}
The peculiarity of this pairing Hamiltonian is that the pairing takes place
within discrete energy levels. This effective Hamiltonian for electrons
pairing due to an attractive interaction brought about by ripplon exchange can
be solved by introducing a variational many body wave function as in the BCS
treatment. However, we chose to apply the Richardson
method~\cite{RichardsonPL3}, initially developed in the context of nuclear
physics and recently reintroduced~\cite{vonDelftSSP39} to the condensed matter
community to describe superconductivity in nanosize metallic grains. This
method is particularly suitable for finite systems with a discrete level
structure such as the multielectron bubble.

\bigskip

\section{Richardson model for pairing in a S2DEG}

Having argued that multielectron bubbles are a suitable candidate to observe
pairing of electrons in a spherical two-dimensional system, we will apply the
Richardson model to the effective pairing Hamiltonian (\ref{Heff}) to gain
insight in the properties of the paired phase.

Note that the analysis here can be applied to the general problem of a S2DEG
with attractive interactions between electrons on the same angular momentum
level. The previous section provides possible values for the coupling constant
$G$ (see Fig. \ref{fig2}) and for the relevant energy scales (see Table I),
and a justification for the applicability of the effective pairing Hamiltonian
(\ref{Heff}) to multielectron bubbles specifically. Nevertheless the results
derived in this section can be used to investigate other systems such as the
superconducting properties of thin electronic nanoshells, or can be
investigated as an academic question regarding spherical electronic systems.

\bigskip

\subsection{Energy levels of interacting electrons}

\bigskip

The Richardson model provides a method of solution for the so-called reduced
BCS Hamiltonian \cite{vonDelftPRL77,MatveevPRL79}:%
\begin{equation}
\hat{H}_{\text{BCS}}=\sum_{i,\sigma}\varepsilon_{i}\hat{b}_{i,\sigma}^{+}%
\hat{b}_{i,\sigma}-G\sum_{i,i^{\prime}}\hat{b}_{i^{\prime},\uparrow}^{+}%
\hat{b}_{i^{\prime},\downarrow}^{+}\hat{b}_{i,\downarrow}\hat{b}_{i,\uparrow}.
\label{HRich}%
\end{equation}
Now consider the Hamiltonian (\ref{Heff}) that we derived in the previous
section, and collect all terms that contain operators working on the angular
momentum level $L$:%
\begin{align}
\hat{H}_{\text{L}}  &  =\sum_{m,\sigma}\epsilon_{L}\hat{c}_{L,m;\sigma}%
^{+}\hat{c}_{L,m;\sigma}\nonumber\\
&  -G\sum_{m,m^{\prime}=-L}^{L}\hat{c}_{L,-m^{\prime}\downarrow}^{+}\hat
{c}_{L,m^{\prime};\uparrow}^{+}\hat{c}_{L,m;\uparrow}\hat{c}_{L,-m;\downarrow
}. \label{HL}%
\end{align}
Note that only electrons within the same angular momentum energy level $L$
interact: only \emph{intra}level interactions take place, but no
\emph{inter}level interactions. This is due to the fact that the relevant
ripplon energies are much smaller than the interlevel energy splitting (see
Table I for typical values in multielectron bubbles). So, the set of electrons
with a given angular momentum $L$ can be considered as an independent
subsystem, described by the Hamiltonian\ (\ref{HL}). The full system is the
collection of independent subsystems characterized by different $L$. The full
Hamiltonian (\ref{Heff}) is just the sum of the Hamiltonians (\ref{HL})
describing independent subsystems with different $L$ :
\begin{equation}
\hat{H}_{\text{eff}}=\sum_{L}\hat{H}_{\text{L}}.
\end{equation}
Moreover, each of the Hamiltonians (\ref{HL}) corresponds to a reduced BCS
Hamiltonian (\ref{HRich}) that can be solved with the Richardson method.
Indeed, setting $\forall i:\varepsilon_{i}=\epsilon_{L}$ and
\begin{equation}
\hat{b}_{m;\uparrow}=\hat{c}_{L,m;\uparrow},\qquad\hat{b}_{m;\downarrow}%
=\hat{c}_{L,-m;\downarrow},
\end{equation}
brings (\ref{HL}) into the same form as (\ref{HRich}). So, we have a
collection of independent systems (each characterized by a particular value of
$L$) that can each be solved by the Richardson method. As shown by
Richardson\ \cite{RichardsonPL3}, the exact solution of the reduced BCS
Hamiltonian for $n$ electron pairs amounts to solving a set of $n$ nonlinear
coupled equations. In general, the aforementioned set of equations can be
solved only by numerical computation. However, in the particular case when all
the involved single-particle states belong to one an the same energy level --
as is the case for a spherical multielectron bubble -- the energy of electron
pairs can be easily found analytically (see, e.g., Ref.
\onlinecite{GladilinPRB70}). The result for the energy of electrons in the
subsystem characterized by angular momentum $L$ can be written down as
\begin{equation}
E_{L,n_{L},g_{L},b_{L}}=(2n_{L}+b_{L})\epsilon_{L}-G(n_{L}-g_{L}%
)(2L-b_{L}+2-n_{L}-g_{L}) \label{EL}%
\end{equation}
The energy levels of the subsystem with angular momentum $L$ is characterized
by three quantum numbers, $n_{L}$, $b_{L}$ and $g_{L}$. Here $n_{L}$ is the
number of electrons pairs, $b_{L}$ is the number of unpaired electrons, and
$g_{L}$ of elementary bosonic pair-hole excitations~\cite{RomanPRB67} in the
system of $n_{L}$ pairs.

To better understand these quantum numbers, consider two bare single-electron
states $\left\vert L,m;\uparrow\right\rangle $ and $\left\vert L,-m;\downarrow
\right\rangle $. If both states are occupied, this represents an electron
pair. The electron pair can scatter into another pair of states $\left\vert
L,m^{\prime};\uparrow\right\rangle ,\left\vert L,-m^{\prime};\downarrow
\right\rangle $ under the influence of the interaction term in the Hamiltonian
(\ref{HL}). The number of such pairs with given $L$ is $n_{L},$ and it has to
be less than or equal to $2L+1$. Now consider the case where only one state of
the pair $\left\vert L,m;\uparrow\right\rangle $ and $\left\vert
L,-m;\downarrow\right\rangle $ is occupied. Then we have an unpaired electron
that cannot participate in the scattering described by the interaction term in
(\ref{HL}). Moreover, electron pairs cannot scatter into the pair of states
$\left\vert L,m;\uparrow\right\rangle $,$\left\vert L,-m;\downarrow
\right\rangle $ because one of these states is already occupied. The states
$\left\vert L,m;\uparrow\right\rangle $,$\left\vert L,-m;\downarrow
\right\rangle $ are then blocked for scattering of pairs. The number of these
blocked spin-degenerate bare states equals the number of unpaired electrons,
and is denoted by $b_{L}$. The total number of electrons in angular momentum
level $L$ is then $2n_{L}+b_{L}\ $and this has to be less than or equal to
$2(2L+1)$. The last quantum number $g_{L}$ indexes the so-called `pair-hole
excitations' \cite{RomanPRB67}. These are bosonic excitations that involve a
redistribution of the amplitude for correlated pairs over the $2L+1$ states
with angular momentum $L$. Note that $g_{L}=0$ corresponds to the ground state
for the correlated pairs.

The total energy can be expressed as a sum of the energies for each
independent subsystem:
\begin{equation}
E_{\{n_{L},b_{L},g_{L}\}_{L=1,2,3,...}}=\sum_{L=0}^{\infty}E_{L,n_{L}%
,g_{L},b_{L}}. \label{Etotal}%
\end{equation}
The state of the entire system is characterized by a large set of quantum
numbers, three ($n_{L},b_{L},g_{L}$) for each angular momentum subsystem.

\bigskip

\subsection{Ground state properties}

\bigskip

How do we characterize the ground state of electrons in a spherical bubble
with pairing interactions ? In bulk BCS superconductors only electrons in an
energy band of the Debye energy $\hbar\omega_{D}$ around the Fermi level
participate in the pairing. However, in the present case, the relevant ripplon
frequencies are much smaller than the splitting between consecutive $L$ levels
near the Fermi energy. Therefore one might argue that pairing correlations
only take place in the one subsystem with $L=L_{F}$ where $L_{F} $ is the
angular momentum at the Fermi level. Indeed, all levels above the Fermi level
($L>L_{F}$) are empty, and all levels below the Fermi level ($L<L_{F}$) are
completely filled so that no pairing correlations can be achieved by
ripplon-mediated scattering.

Yet this turns out to be wrong. The main difference between pairing in the
present case and pairing in conventional superconductors is that in
conventional BCS superconductors the gap $\Delta$ is much smaller than the
relevant phonon energy ($\Delta\ll\hbar\omega_{D}$), whereas in the present
case the interaction energy per electron can be larger than the relevant
ripplon energies ($\Delta\gg\hbar\omega_{r}$) and even larger than the level
splitting ($\Delta>\epsilon_{L_{F}}-\epsilon_{L_{F}-1}$). This can be inferred
by comparing the interaction strengths listed in Table II to the ripplon
energies and level splittings reported in Table I. It then becomes
energetically advantageous to redistribute the electrons amongst the energy
levels $L$. Indeed, by promoting a pair of electrons from a level $L_{1}$
below the Fermi level $(L_{1}<L_{F})$ to a level $L_{2}$ above the Fermi level
($L_{2}>L_{F}$), both these levels $L_{1},L_{2}$ can also form pairing
correlations. The energy gain by forming pairing correlations is larger than
the energy needed to promote the electrons form level $L_{1}$ to level $L_{2}
$. So we obtain the remarkable result that for sufficiently large interaction
strength $G$ (as we think is the case for MEBs), although only intralevel
scattering can take place, still levels well below and above the Fermi level
can be affected.

\bigskip

To see this in more detail, let us consider for example a bubble with an even
number of electrons. The ground state in an even bubble is achieved by
$g_{L}=0$ (no pair-hole excitations) and $b_{L}=0$ (no unpaired electrons). In
order to find the ground-state configuration $n_{L}$, consider the change in
the total energy due to a transfer of an electron pair from the $L^{\text{th}%
}$ level to the next higher level. Using Eq. (\ref{EL}) with $g_{L}=0$ and
$b_{L}=0$, we find%
\begin{equation}
\Delta E_{L,L+1}=2\epsilon_{1}(L+1)-2G(n_{L}-n_{L+1}) \label{dEtransfer}%
\end{equation}
where $n_{L}$ and $n_{L+1}$ are the number of electron pairs on the levels $L
$ and $L+1$, respectively, before the transfer. The first term in the
right-hand side of (\ref{dEtransfer}) represents the energy cost in promoting
a pair of electrons from level $L$ to level $L+1$, and the second term
corresponds to the gain in energy due to pairing correlations. As seen from
(\ref{dEtransfer}), such a transfer reduces the total energy if the
inequality
\begin{equation}
G(n_{L}-n_{L+1})>\epsilon_{1}(L+1) \label{condtransf}%
\end{equation}
is satisfied. This inequality is not satisfied if $n_{L}<n_{L+1}$, so that an
inverted population of the levels obviously never appears in the ground state.
Condition (\ref{condtransf}) is also not fulfilled at weak interaction, i.e.
for $G/\epsilon_{1}<(L+1)/(n_{L}-n_{L+1})$. In the case of $L\gg1,$ the above
definition of weak interaction can be simplified to $G<\hbar^{2}/(2m_{e}%
R^{2})$. From Table I and Table II, we infer that for MEBs it is likely that
$G>\hbar^{2}/(2m_{e}R^{2})$ so that a redistribution of electron pairs as
described in the previous paragraph is indeed energetically favorable.

\bigskip

\begin{figure}
[ptb]
\begin{center}
\includegraphics[
height=2.983in,
width=5.3931in
]%
{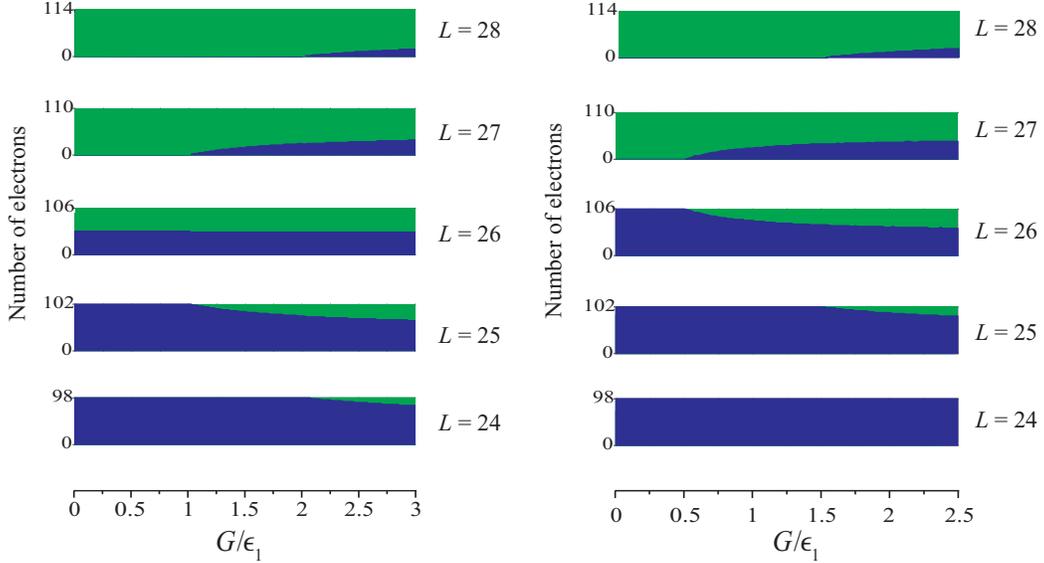}%
\caption{Filled and empty states are indicated with blue resp. green and shown
as a function of the dimensionless interaction strength $G/\epsilon_{1}.$ The
typical ripplon energies are smaller than the splitting between successive $L$
levels (with degeneracy $2L+1$), so that only intralevel scattering takes
splace. Nevertheless the electrons are redistributed over different $L$ levels
around the Fermi level $L_{F}=26$, because the pairing energy is comparable to
the level splitting. The left panel shows the result for $N_{F}=52$ and the
right panel for $N_{F}=106$ electrons on the Fermi level at $G=0$.}%
\label{fig3}%
\end{center}
\end{figure}

Fig. \ref{fig3} illustrates the ground-state configuration obtained by
minimizing the total energy (\ref{Etotal}) with respect to the $n_{L}$'s (and
setting $b_{L}=g_{L}=0$). The results are shown as a function of
$G/\epsilon_{1}$ for fixed $L_{F}=26$. The left panel shows the result for
$N_{F}=52$ electrons on the Fermi level in the ground state of an MEB with
$G=0.$ This corresponds to roughly half-filling of the Fermi level. The right
panel shows the result for $N_{F}=106$, a closed-shell configuration in the
$G=0$ ground state. Blue (green) color corresponds to filled (empty) states on
the energy levels. Switching off the interactions, $G=0$, we see that all
levels below $L_{F}=26$ are completely filled (blue color), and all levels
above $L_{F}$ are completely empty (green color). For $G/\epsilon_{1}=2.5$,
some empty states (green) appear in levels $L=24,25$ that are below $L_{F}$,
and some electron pairs appear in levels $L=27,28$ that are above $L_{F}$.
Note that $G$ needs to exceed a critical value (of the order of $\epsilon
_{1}/2$) for the redistribution of electron pairs to take place. Also note
that approximately $2G/\epsilon_{1}$ levels around the Fermi level are
affected by the redistribution of electron pairs.%

\begin{figure}
[ptb]
\begin{center}
\includegraphics[
height=2.2325in,
width=2.5695in
]%
{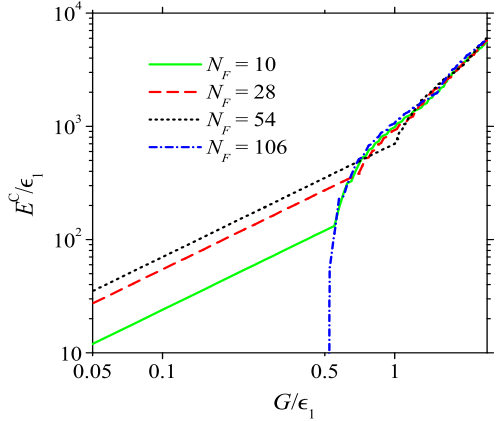}%
\caption{The condensation energy $E^{\text{C}}$ is shown as a function of the
interaction strength $G$, for different numbers of electrons in the Fermi
level $L_{F}=26$ at $G=0$. As the coupling increases, the condensation energy
becomes less sensitive to the precise filling of the Fermi level.}%
\label{fig4}%
\end{center}
\end{figure}

Having obtained, for the ground state, the total energy $E_{\mathrm{g.s.}}$,
we can derive the condensation energy
\begin{equation}
E^{C}=E_{\mathrm{g.s.}}(0)-E_{\mathrm{g.s.}}(G)-G\text{ }\left\lceil
N/2\right\rceil ,
\end{equation}
where $E_{\mathrm{g.s.}}(G)$ [$E_{\mathrm{g.s.}}(0)$] is the ground-state
energy in the presence [in the absence] of the pairing interaction. The last
term in the rhs describes the interaction energy for the uncorrelated Fermi
ground state and $\left\lceil x\right\rceil $ means the integer part of $x$.
In Fig. \ref{fig4} we plot the condensation energy $E^{C}$ as a function of
$G/\epsilon_{1}$. First consider the regime of strong interactions,
$G\gg\epsilon_{1}/2$. For this regime, we find that the $N_{F}-$dependence of
the condensation energy becomes negligible. The condensation energy rapidly
rises with $G/\epsilon_{1}$, approximately as $E^{C}\approx(2L_{F}%
G)^{2}/(3\epsilon_{1})$. Next, consider the regime of weak interactions
($G<\epsilon_{1}/2$) where, as we have seen, the electron redistribution
between the energy levels is absent. In this case the condensation energy is a
linear function of $G$, and such a linear behavior can indeed be seen in Fig.
\ref{fig4} at $G<\epsilon_{1}/2$. In the regime of weak interaction, the
condensation energy is strongly influenced by $N_{F} $: it is zero for closed
shell configurations and reaches a maximum at half filling. For typical MEBs,
we find that $E^{C}/k_{B}$ is of the order of several Kelvin. Decreasing the
number of electrons or pressurizing the bubble increases the condensation
energy. Bigger bubbles have a smaller condensation energy.%

\bigskip

\subsection{Experimental signatures}

\bigskip

How can the pairing correlations be probed experimentally ? What properties
will distinguish the state with pair correlations from the normal Fermi sea ?
In the preceding subsection, we have found the ground state for the system
described by the Hamiltonian $H_{\text{eff}}$, expression (\ref{Heff}), and in
the preceding section we argued, based on Cooper's argument, why this
Hamiltonian is suitable to describe MEBs. Nevertheless, more exotic correlated
or magnetic states not described by the Hamiltonian (\ref{Heff}) cannot be
completely excluded as candidates for the true ground state: only experiment
will give a decisive answer as to what the realized state in the MEB will be.
It is therefore very relevant to discuss accessible experimental signatures of
the correlated many-body state that we have described.

In metallic nanograins and nanowires, pairing correlations (leading to
superconductivity) were revealed through a measurement of the density of
states \cite{BlackPRL76}. Also here, we propose to reveal pairing correlations
by probing the density of states. In the case of stabilized MEBs
\cite{SilveraBAPS}, the density of states can be measured by spectroscopy.
Also tunneling experiments are possible by placing an electrode close to the
bubble, and reducing the thickness of the layer of helium between the bubble
and the helium. The density of states that would be revealed by these
experiments is the subject of the next subsection.

\bigskip

\subsection{Density of states}

\bigskip

In the previous subsection, we found that the ground state is characterized by
a redistribution of electrons between degenerate bare energy levels, as
compared to the ground-state electron distribution in the absence of
interaction. Here we analyze the effect of the pairing interaction on the
density of (many-electron) states in MEBs. This density of states can be
written down in general as
\begin{equation}
\mathcal{D}(E)=\sum_{i}J^{(i)}\delta\left(  E-E_{i}\right)  ,
\end{equation}
From expression (\ref{Etotal}) it is clear that the many-electron energy
levels $E_{i}$ are characterized by a set of quantum numbers \textquotedblleft%
$i$\textquotedblright\ $:=\{n_{L},b_{L},g_{L}\}_{L=1,2,3,...}$. The summation
runs over all possible sets $i$ of quantum numbers. The degeneracy of the
many-electron energy level $E_{i}$ is denoted by $J^{(i)}$. Consider one
particular angular momentum state $L$ with a given \{$n_{L},b_{L},g_{L}$\}.
The number of ways to place the unpaired electrons, multiplied by the number
of pair-hole excitations with given $g_{L}$, is $J_{n_{L},g_{L},b_{L}}$ (see
Refs. \cite{GladilinPRB70,RomanPRB67}) :%
\begin{equation}
J_{n_{L},g_{L},b_{L}}=2^{b_{L}}C_{b_{L}}^{2L+1}\times\left\{
\begin{array}
[c]{l}%
1,\quad g_{L}=0\\
\left(  C_{g_{L}}^{2L+1-b_{L}}-C_{g_{L}-1}^{2L+1-b_{L}}\right)  ,\quad
g_{L}\geq1
\end{array}
\right.
\end{equation}
with $C_{j}^{k}$, the binomial coefficients. The total degeneracy is the
product of the degeneracies of the independent systems,
\begin{equation}
J^{(i)}=\prod_{L}J_{n_{L},g_{L},b_{L}}.
\end{equation}
For graphical representation, it is more convenient to consider instead of
$\mathcal{D}(E)$ the quantity
\begin{equation}
\mathcal{D}_{\delta}(E)=\int_{E-\delta/2}^{E+\delta/2}dE\,\mathcal{D}(E),
\end{equation}
which gives the number of (many-electron) states in the energy range of width
$\delta$ around the energy $E$.

\bigskip

\begin{figure}
[ptb]
\begin{center}
\includegraphics[
height=2.7455in,
width=3.5392in
]%
{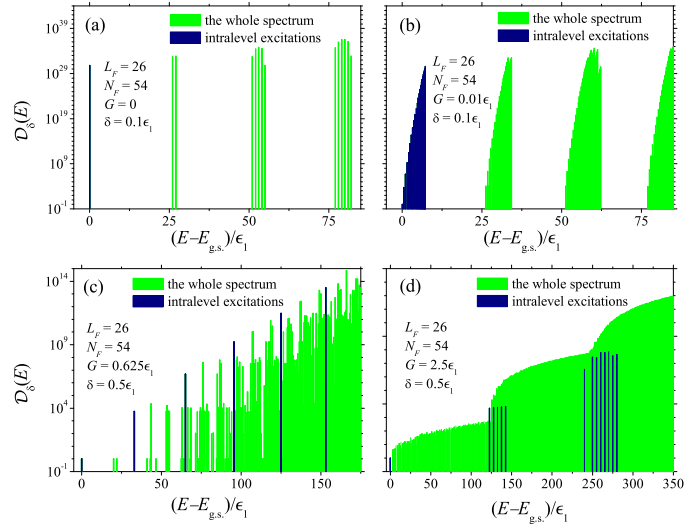}%
\caption{The calculated density of states for a spherical 2D electron system
with pairing, obtained within the Richardson model, is shown in these figures
as a function of the energy above the ground state energy. The different
panels correspond to different strengths of the attractive electron-electron
interaction term in the reduced BCS Hamiltonian.}%
\label{fig5}%
\end{center}
\end{figure}

Fig.~\ref{fig5} gives an example of the calculated $\mathcal{D}_{\delta}(E)$.
The calculations are performed for $N_{F}=54$ electrons in the Fermi angular
momentum level $L_{F}=26$, so that in the ground state at $G=0$ the Fermi
level is approximately half filled. Together with the whole spectrum of
excitations (displayed in light green) we also show (in dark blue) the
intralevel excitations from the ground state.%

At $G=0$, as shown in Fig. \ref{fig5}a, there is a set of excitations that
correspond to different interlevel electron transitions between the
single-electron bare energy levels with $L$ close to $L_{F}$ (obviously,
intralevel transitions have zero energy in this case). At $G=0$, all the
many-electron energy levels in the system under consideration are highly
degenerate. This is due to the numerous possibilities to distribute electrons
over single-electron states belonging to a partially filled energy level with
a given $L$.

At small nonzero $G$, the many-electron energy levels for each angular
momentum level are split.\ Excitations involving pair breaking $(b_{L}\neq0)$
and pair-hole excitations $(g_{L}\neq0)$ start to affect the density of
states. The result, shown in Fig. \ref{fig5}b, is the appearance of energy
bands in the density of states. The first band corresponds to intralevel
excitations from the ground state, the other bands involve interlevel
excitations from one angular momentum state to another.

At this point it is interesting to investigate the gap in the excitation
spectrum for intralevel excitations. In superconducting nanograins, measuring
such a spectroscopic gap signals the onset of superconductivity
\cite{BlackPRL76}. First, consider a process where the ground state (with
$n_{L}$ pairs and $b_{L}=0$ unpaired electrons) is transformed into a final
state with $n_{L}^{\prime}=n_{L}-1$ and $b_{L}^{\prime}=2$ unpaired electrons.
We will refer to such a process as a `pair-breaking excitation'. A
pair-breaking `gap' $\Delta^{(\text{p-b})}$ can then be defined as the energy
necessary to create a pair-breaking excitation from the ground state:%
\begin{equation}
\Delta_{L}^{(\text{p-b})}=E_{L,n_{L}-1,0,2}-E_{L,n_{L},0,0}.
\end{equation}
Using (\ref{EL}),
\begin{equation}
\Delta_{L}^{(\text{p-b})}=G(2L+1).
\end{equation}
Similarly, we can calculate the smallest energy needed to create a pair-hole
excitation ($g_{L}=0\rightarrow g_{L}=1$) and define a pair-hole excitation
gap $\Delta^{(\text{p-h})}$%
\begin{align}
\Delta_{L}^{(\text{p-h})}  &  =E_{L,n_{L},1,0}-E_{L,n_{L},0,0}\nonumber\\
&  =G(2L+1).
\end{align}
So, in the system under consideration, the smallest energy for a pair-breaking
excitation equals that for a pair-hole excitation, $\Delta^{(\text{p-b}%
)}=\Delta^{(\text{p-h})}$, and we can drop the superscripts (p-b) and (p-h).
At $G\approx\epsilon_{1}/2$ the intralevel excitation gap $\Delta$ approaches
the energy spacing $\epsilon_{1}(L+1)$ between single-electron bare energy
levels with consecutive values of $L$. This means that, for $G\gg\epsilon
_{1}/2$, interlevel excitations exist with energy smaller than the gap
$\Delta.$ Moreover, as implied by Eq.~(\ref{dEtransfer}), an increase of $G$
can substantially reduce the energies of interlevel transitions of pairs. At
$G=0.625\epsilon_{1}$ (see Fig. \ref{fig5}c), the lowest interlevel
excitations (green) already have energies smaller than $\Delta\ $(first
nonzero blue line).

At even larger $G$, pairing correlations appear on many different angular
momentum levels. The pair-breaking/pair-hole excitation gap, $\Delta_{L}$,
depends on the angular momentum. Therefore, as seen from Fig. \ref{fig4}d, the
peaks of $\mathcal{D}_{\delta}(E)$ corresponding to intralevel excitations
from the ground state, are split into peaks corresponding to different angular
momentum states where pairing can take place. With increasing $G$, the
excitation spectrum tends to become quasi-continuous, with jumps of several
orders of magnitude in $\mathcal{D}_{\delta}(E)$ near the energies
corresponding to pair-breaking or pair-hole excitations. Between these jumps
there is a more uniform distribution of excitations, corresponding to
interlevel transitions of pairs.

\begin{figure}
[ptb]
\begin{center}
\includegraphics[
natheight=9.080100in,
natwidth=5.855500in,
height=4.4948in,
width=2.9091in
]%
{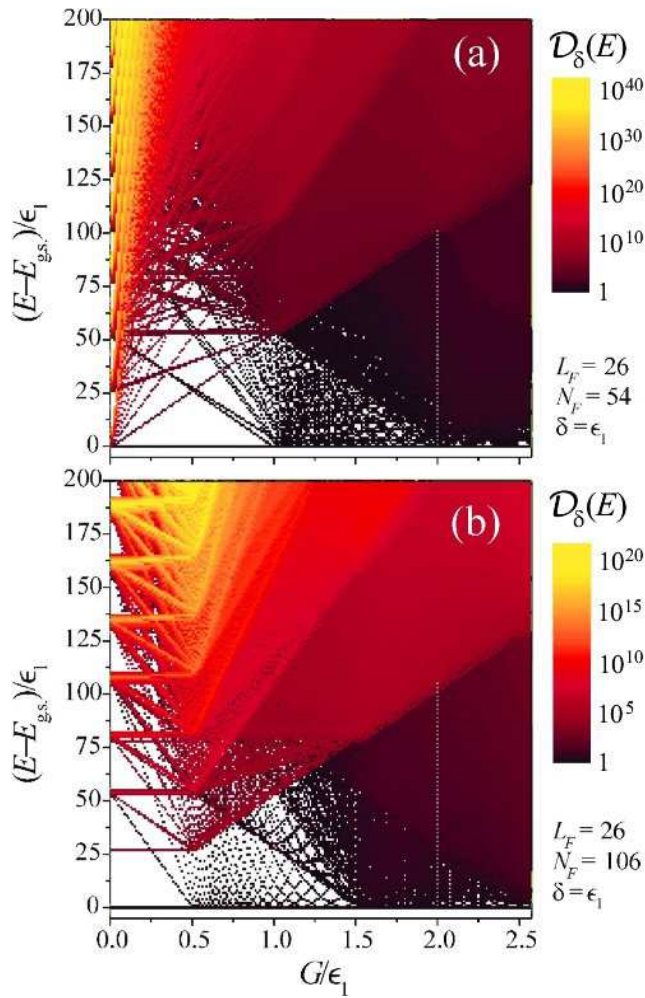}%
\caption{Density of states of the spherical 2D electron system with pairing
interactions, as a function of the interaction strength and the energy above
the ground state, both expressed in energy units $\epsilon_{1}$. The results
are shown for two different values of $N_{F}$, the number of electrons on the
Fermi level at $G=0$: (a) $N_{F}=54$ (approximately half filling) and (b)
$N_{F}=106$ (closed-shell configuration).}%
\label{fig6}%
\end{center}
\end{figure}

Fig. \ref{fig6} provides an \textquotedblleft overview\textquotedblright\ of
the behavior of $\mathcal{D}_{\delta}(E)$ as a function of both $E$ and $G$
for fixed $L_{F}=26$ and two different values of $N_{F}$. One can see an
interplay between intralevel excitations, whose energies always increase with
increasing $G$, and excitations corresponding to interlevel transitions of
pairs. For the latter, both an increase and a decrease in energy are possible
with increasing $G$. While at $G\lesssim\epsilon_{1}$ the patterns of
$\mathcal{D}_{\delta}(E,G)$ are very different for different $N_{F}$, at
larger $G$ the behavior of $\mathcal{D}_{\delta}(E)$ in MEBs with a definite
parity of the number of electrons becomes almost independent of the precise
value of this number [cf. Figs.~\ref{fig6}(a) and~\ref{fig6}(b)]. This
\textquotedblleft universal behavior\textquotedblright\ of $\mathcal{D}%
_{\delta}(E)$ at large $G/\epsilon_{1}$ is further illustrated by
Fig.~\ref{fig6}, where the calculated dependence of $\mathcal{D}_{\delta}$ on
$(E-E_{\mathrm{g.s.}})/\epsilon_{1}$ is shown for different $N_{F}$ and
$L_{F}$. As seen from Fig.~\ref{fig6}, neither moderate changes of $L_{F}$ nor
variations of $N_{F}$ at a fixed $L_{F}$ significantly affect the shape of
$\mathcal{D}_{\delta}$ versus $(E-E_{\mathrm{g.s.}})/\epsilon_{1}$ for MEBs
with a given parity of the number of electrons. At the same time,
Fig.~\ref{fig6} demonstrates a pronounced difference between the results for
even MEBs and those for odd MEBs.%

%

\begin{figure}
[ptb]
\begin{center}
\includegraphics[
height=3.8904in,
width=2.8177in
]%
{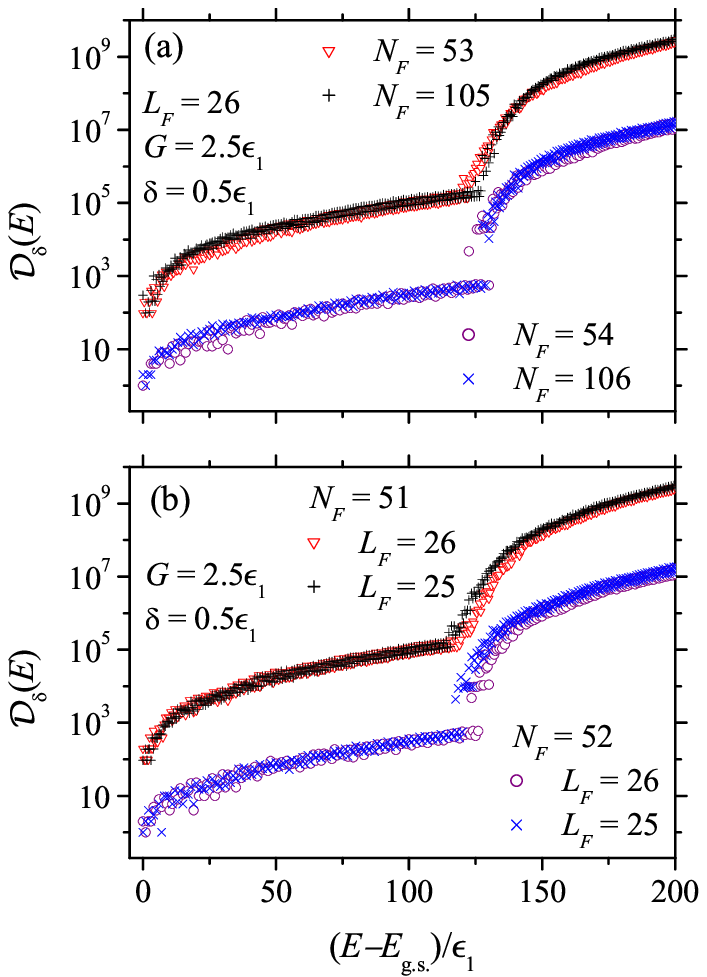}%
\caption{Density of states as a function of the energy above the ground state
energy for MEBs with the pairing-interaction strength $G=2\epsilon_{1}$ and
different number of electrons. Panel (a) shows the results for $L_{F}=26$ and
different values of $N_{F}$, which determines the filling of the Fermi level
at $G=0$. Panel (b) shows the results for two different $L_{F}$ and different
parity of the number of electrons in an MEB in the case when the Fermi level
is approximately half-filled at $G=0$.}%
\label{fig7}%
\end{center}
\end{figure}

Summarizing this subsection, we find that in the regime most relevant for
multielectron bubbles, namely $G\gg\epsilon_{1}$, the pairing correlations
reveal themselves in the density of states not as a spectroscopic gap, but
rather as a significant jump in the density of states at the pair-breaking
energy $\Delta$. Moreover, in this regime ($G\gg\epsilon_{1}$) the density of
states is less sensitive to even-odd effects. The observation (through
tunneling or spectroscopy) of a jump in the density of states at $\Delta$ can
be used as a way to infer the presence of pairing correlations in the MEB.
Moreover, the pair-breaking energy can be used to estimate a temperature
$T_{c}$ above which the pairing correlations will be suppressed: when
$k_{B}T>\Delta$ the thermal energy is large enough to support an appreciable
amount of pair-breaking excitations. For typical MEBs, this temperature is of
the order of Kelvins. Smaller bubbles or compressed bubbles have a larger gap
and thus a larger $T_{c}$.

\bigskip

\section{Conclusions}

\bigskip

In the first part of this paper, we have analyzed how the electron-ripplon
interaction on a spherical surface may lead to an attractive effective
electron-electron interaction and give rise to a Cooper pairing scenario. The
effective Hamiltonian of the two dimensional spherical electron system is
mapped on a BCS-type Hamiltonian and typical values of energies, length scales
and interaction strengths are estimated.

In the second part we use Richardson's method to investigate pairing
properties of a two dimensional spherical electron system. We find that when
the condensation energy per pair is larger than the bubble energy scale
$\hbar^{2}/(m_{e}R^{2}),$ the ground state of the system acquires unique
properties that set it apart from pairing in conventional superconductors or
superconducting nanograins. In particular, we show that although only
intralevel interactions are included (since the relevant ripplon energies are
smaller than the level splitting), electron pairs nevertheless redistribute
themselves among the different levels and pairing takes place in an interval
of energies around the Fermi energy, much larger than the typical ripplon
energy. The density of states reveals an intricate interplay between
intralevel transitions and interlevel excitation of pairs, evolving from a
discrete spectrum typical for confined systems to a staircase quasi-continuum
upon increasing interaction strength. At strong coupling the density of states
reveal the presence of pairing correlations not through a spectroscopic gap as
in metallic nanograins \cite{BlackPRL76}, but through a jump in the density of
states at the pair-breaking energy.

These results show that spherical electron systems reveal particularly
interesting pairing properties, distinct from their bulk or flat-surface
counterparts, combining both topological effects and confinement effects. MEBs
are a particularly pure realization of the two-dimensional electron system
(just as an electron film on helium forms a pure realization of a flat 2DEG).
Thin nanoshells (monolayer gold coatings of a non-conducting nanograin) may be
another realization of this system.

\bigskip

\begin{acknowledgments}
J.T. is supported financially by the Fonds voor Wetenschappelijk Onderzoek -
Vlaanderen. This research has been supported financially by the FWO-V projects
Nos. G.0435.03, G.0306.00, G.0274.01N, the W.O.G. project WO.035.04N, the GOA
BOF UA 2000. Also financial support from the Department of Energy, Grant No.
DE-FG02-ER45978 is acknowledged. J.T. gratefully acknowledges support of the
Special Research Fund of the University of Antwerp, BOF\ NOI UA 2004.
\end{acknowledgments}

\end{document}